# Crystal Structure and Superconductivity at about 30 K in $A$Ca$_2$Fe$_4$As$_4$F$_2$ ($A$ = Rb, Cs)

Zhi-Cheng Wang[1], Chao-Yang He[1], Zhang-Tu Tang[1], Si-Qi Wu[1] and Guang-Han Cao[1,2]*

## Abstract

We have synthesized two iron fluo-arsenides $A$Ca$_2$Fe$_4$As$_4$F$_2$ with $A$ = Rb and Cs, analogous to the newly discovered superconductor KCa$_2$Fe$_4$As$_4$F$_2$. The quinary inorganic compounds crystallize in a body-centered tetragonal lattice with space group I4/mmm, which contain double Fe$_2$As$_2$ layers that are separated by insulating Ca$_2$F$_2$ layers. Our electrical and magnetic measurements on the polycrystalline samples demonstrate that the new materials undergo superconducting transitions at $T_c$ = 30.5 K and 28.2 K, respectively, without extrinsic doping. The correlations between $T_c$ and structural parameters are discussed.

[1] Department of Physics and Stat Key Lab of Silicon Materials, Zhejiang University, Hangzhou 310027, China
[2] Collaborative Innovation Centre of Advanced Microstructures, Nanjing 210093, China
* Corresponding author (email: ghcao@zju.edu.cn)

Since the discovery of high-temperature superconductivity in iron pnictides [1,2], considerable efforts have been devoted to explore new iron-based superconductors (IBS). To date dozens of IBS, in diverse structure types, have been discovered, all of which contain $Fe_2X_2$ ($X$ = As or Se) layers that are essential for the emergence of superconductivity. Some of the structures are relatively simple because fewer atoms connect the superconductive $Fe_2X_2$ layers [1-7]. Other complex structures consist of thicker connecting blocks between the $Fe_2X_2$ layers [8-13]. These material discoveries greatly enrich the related research on IBS [14-16].

In 2013 we designed several new structures containing $Fe_2X_2$ layers, in view of the crystal chemistry of IBS [17]. This year one of the candidate structure "KLaFe$_4$As$_4$" was realized in the form of $AkAe$Fe$_4$As$_4$ ($Ak$ = K, Rb, Cs; $Ae$ = Ca, Sr) [18] and $Ak$EuFe$_4$As$_4$ ($Ak$ = Rb, Cs) [19-21]. Very recently, another proposed structure $A_3Fe_4X_4Z_2$ (see Ref. 17 for details of the definition of $A$, $X$ and $Z$) was implemented in KCa$_2$Fe$_4$As$_4$F$_2$ (hereafter called 12442) [22]. In these new IBS above, the $Fe_2X_2$ layers are asymmetric, not being seen in previous IBS. Furthermore, these materials are self doped (hence superconductivity emerges) in their stoichiometric form. For the 12442-type IBS, in particular, there are double Fe$_2$As$_2$ layers (with alkali metal atoms sandwiched) separated by Ca$_2$F$_2$ layers, which has not been seen in previous IBS either.

In this work, we synthesized $A$Ca$_2$Fe$_4$As$_4$F$_2$ ($A$ = Rb, Cs), two sister compounds of KCa$_2$Fe$_4$As$_4$F$_2$, which can be viewed as an intergrowth of $A$Fe$_2$As$_2$ ($A$ = Rb, Cs) and CaFeAsF. The crystallographic parameters were determined. The physical property measurements demonstrate that RbCa$_2$Fe$_4$As$_4$F$_2$ and CsCa$_2$Fe$_4$As$_4$F$_2$ exhibit bulk superconductivity at 30.5 K and 28.2 K, respectively.

Polycrystalline samples of RbCa$_2$Fe$_4$As$_4$F$_2$ and CsCa$_2$Fe$_4$As$_4$F$_2$ were prepared by conventional solid-state reactions in sealed Ta tubes. Details of the sample preparation are presented in the following **Experimental Section**.

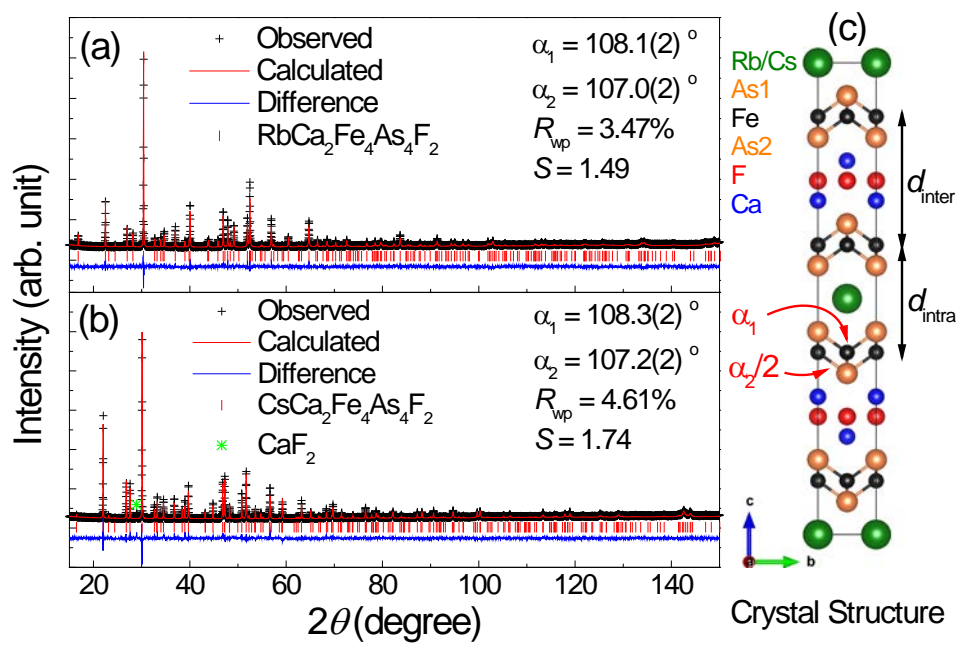

**Figure 1** X-ray diffraction patterns and their Rietveld refinement profiles for RbCa$_2$Fe$_4$As$_4$F$_2$ (a) and CsCa$_2$Fe$_4$As$_4$F$_2$ (b) polycrystalline samples. (c) The 12442-type crystal structure, projected along the [100] direction.

Powder X-ray diffraction (XRD) patterns of as-prepared RbCa$_2$Fe$_4$As$_4$F$_2$ and CsCa$_2$Fe$_4$As$_4$F$_2$ are displayed in Fig. 1(a) and Fig. 1(b), respectively. The former shows no obvious impurities, and the latter indicates nearly pure 12442-type phase with weak reflections (3% of the strongest reflection) from unreacted CaF$_2$. These XRD data were then employed for Rietveld analyses using software Rietan-FP [23]. The refinement was based on the 12442-type structural model shown Fig. 1(c) in which the occupation factor of each atom was fixed to 1.0. The weighted reliability factor $R_{wp}$ are 3.47% and 4.61%, while the goodness-of-fit parameter $S$ are 1.49 and 1.74, for RbCa$_2$Fe$_4$As$_4$F$_2$ and CsCa$_2$Fe$_4$As$_4$F$_2$ respectively. The resulting crystallographic data are tabulated in Table 1 in comparison with those of KCa$_2$Fe$_4$As$_4$F$_2$ [22].

**Table 1** Crystallographic data of $A$Ca$_2$Fe$_4$As$_4$F$_2$ ($A$ = K, Rb, Cs). The space group is I4/mmm (No. 139). The atomic coordinates are as follows: K/Rb/Cs 2$a$ (0, 0, 0); Ca 4$g$ (0.5, 0.5, $z$); Fe 8$g$ (0.5, 0, $z$); As1 4$e$ (0.5, 0.5, $z$); As2 4$e$ (0, 0, $z$); F 4$d$ (0.5, 0, 0.25). $\alpha_1$ and $\alpha_1$ denote the angles of As1−Fe−As1 and As2−Fe−As2, respectively.

| $A$ | K (Ref. 22) | Rb | Cs |
|---|---|---|---|
| **Lattice Parameters** | | | |
| $a$ (Å) | 3.8684(2) | 3.8716(1) | 3.8807(1) |
| $c$ (Å) | 31.007(1) | 31.667(1) | 32.363(1) |
| $V$ (Å$^3$) | 463.99(3) | 474.66(1) | 487.38(1) |
| **Coordinates ($z$)** | | | |
| Ca | 0.2085(2) | 0.2089(2) | 0.2100(2) |
| Fe | 0.1108(1) | 0.1140(1) | 0.1172(1) |
| As1 | 0.0655(1) | 0.0697(1) | 0.0739(1) |
| As2 | 0.1571(1) | 0.1592(1) | 0.1614(1) |
| **Bond distances (Å)** | | | |
| Fe−As1 | 2.390(3) | 2.391(2) | 2.393(2) |
| Fe−As2 | 2.409(3) | 2.407(2) | 2.410(2) |
| **As Height from Fe Plane (Å)** | | | |
| As1 | 1.405(3) | 1.403(1) | 1.401(1) |
| As2 | 1.436(3) | 1.431(1) | 1.430(1) |
| **Bond Angles (°)** | | | |
| $\alpha_1$ | 108.0(2) | 108.1(2) | 108.3(2) |
| $\alpha_2$ | 106.8(2) | 107.0(2) | 107.2(2) |

As an intergrowth between RbFe$_2$As$_2$ and CaFeAsF, RbCa$_2$Fe$_4$As$_4$F$_2$ indeed shows a lattice constant $a$ very close to the mean value of those of RbFe$_2$As$_2$ (3.863 Å) [24] and CaFeAsF (3.878 Å) [25]. Meanwhile the $c$ axis is almost equal to the expected value (2$c_{CaFeAsF}$ + $c_{RbFe2As2}$ = 31.633 Å). Similarly, the lattice constants of CsCa$_2$Fe$_4$As$_4$F$_2$ are related to those of CaFeAsF and CsFe$_2$As$_2$ [26]. Note that the apparent Fe valence in the hybrid 12442 phase become 2.25+, an average of 2.5+ (in RbFe$_2$As$_2$) and 2+ (in CaFeAsF). Along with charge homogenization, the constituent building blocks in the 12442-type structure suffer subtle modifications. For example, the CaFeAsF block becomes slightly slimmer, as is also seen in KCa$_2$Fe$_4$As$_4$F$_2$ [22].

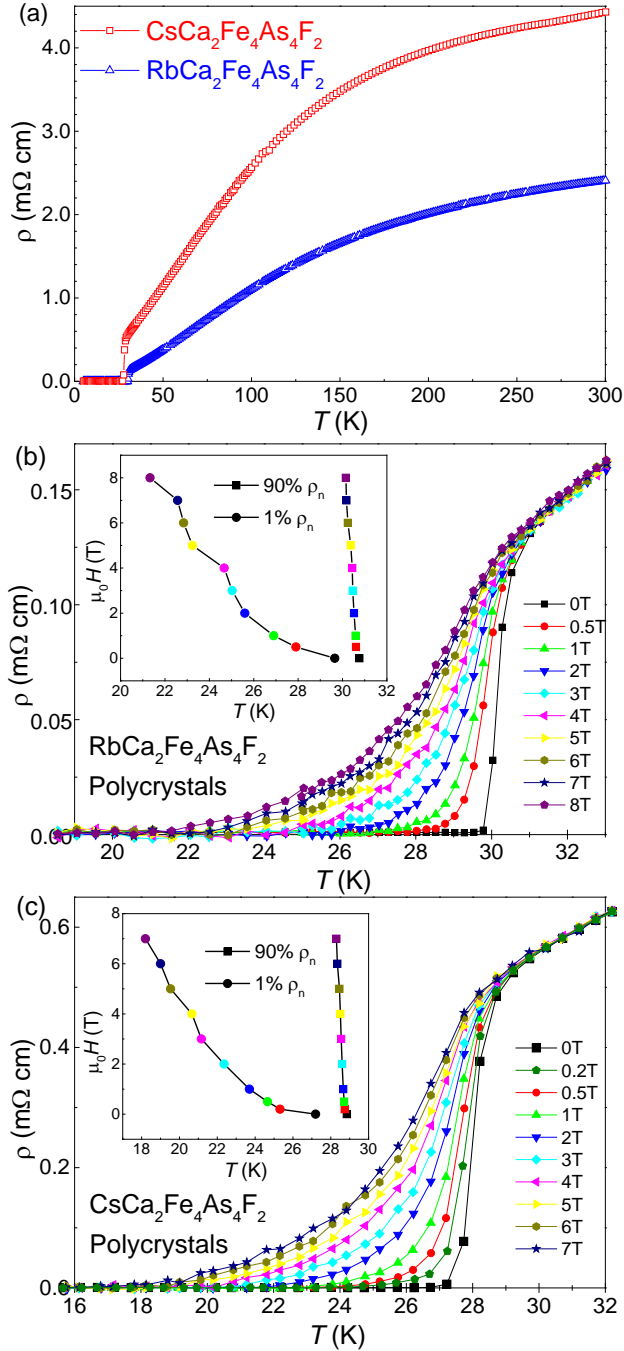

**Figure 2** Temperature dependence of electrical resistivity for the RbCa$_2$Fe$_4$As$_4$F$_2$ and CsCa$_2$Fe$_4$As$_4$F$_2$ polycrystalline samples. The insets of (b) & (c) are derived upper critical field ($H_{c2}$) and irreversible field ($H_{irr}$) as functions of temperature.

Fig. 2(a) shows temperature dependence of electrical resistivity [$\rho(T)$] for $A$Ca$_2$Fe$_4$As$_4$F$_2$ ($A$ = Rb, Cs) samples. Overall, the $\rho$(T) curves show a metallic behavior. No anomaly associated with the spin-density wave ordering that often appears in the parent compounds of IBS [14-16] can be seen, consistent with the self hole doping at 25%. Nevertheless, there is a hump-like anomaly at around 150 K, which is commonly seen in hole-doped IBS [3,27]. Notably, the resistivity decreases almost linearly below ~75 K until a superconducting transition takes place. The onset superconducting transition temperature ($T_c^{onset}$) is defined as the interception of linear extrapolations from superconducting and normal-state sides. For $A$ = Rb and Cs, $T_c^{onset}$ are 30.5 K and 28.5 K, respectively, which are somewhat lower than that of $KCa_2Fe_4As_4F_2$ (33.0 K) [22].

Upon applying magnetic field, the superconducting transitions shift to lower temperatures with a pronounced tail, as shown in Fig. 2(b) & (c). Consequently, the zero-resistance temperature $T_c^{zero}$ decreases much faster than $T_c^{onset}$, suggesting high anisotropy in the 12442-type materials. Taken 90% and 1% of $\rho_n$ (the extrapolated normal-state value at $T_c$) as the criteria, the upper critical field ($H_{c2}$) and the irreversible field ($H_{irr}$) are extracted, both of which are plotted in the insets of Fig. 2(b) & (c). The large gaps between $H_{c2}(T)$ and $H_{irr}(T)$ confirms the enhanced two-dimensionality. Additionally, the slopes of $H_{c2}(T)$ for $RbCa_2Fe_4As_4F_2$ and $CsCa_2Fe_4As_4F_2$ achieve as high as 13.9 T/K and 14.2 T/K, respectively, suggesting small superconducting coherence lengths at zero temperature. Further measurements using single crystalline samples are needed to reveal the anisotropy directly.

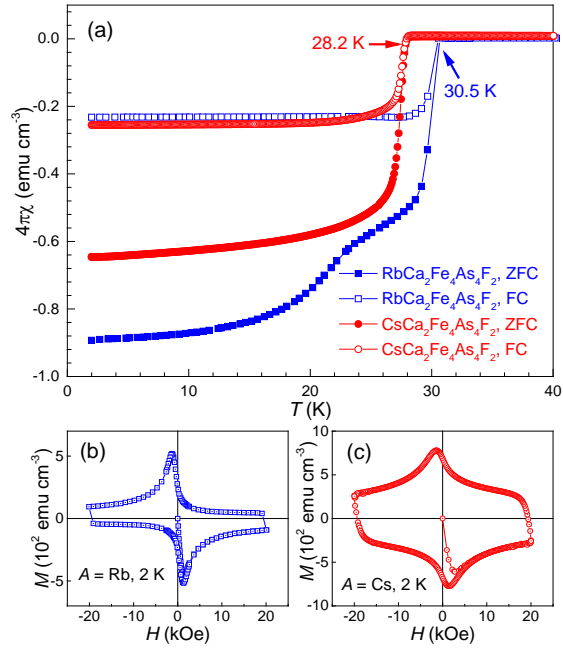

**Figure 3** (a) Temperature dependence of magnetic susceptibility measured at 10 Oe in field-cooling (FC) and zero-field-cooling (ZFC) modes for $ACa_2Fe_4As_4F_2$ ($A$ = Rb, Cs). (b) & (c) Isothermal magnetization loops at 2 K for $RbCa_2Fe_4As_4F_2$ and $CsCa_2Fe_4As_4F_2$.

Fig. 3(a) shows temperature dependence of dc magnetic susceptibility measured under a low field of 10 Oe. The diamagnetic transitions occur at 30.5 K and 28.2 K, consistent with the resistivity measurement. After a demagnetization correction, the volume fractions of magnetic shielding, measured in zero-field-cooling (ZFC) mode, reach 90% and 65%, respectively, for $RbCa_2Fe_4As_4F_2$ and $CsCa_2Fe_4As_4F_2$. The Meissner volume fractions, measured in field-cooling (FC) mode, still achieve 23% and 25%, which unambiguously manifest bulk superconductivity. Fig. 3(b) & (c) show the isothermal magnetization curves at 2 K for the two superconductors. The lower critical magnetic fields $H_{c1}$ (about 400 Oe) are much smaller than the upper critical magnetic fields (see above), therefore, $ACa_2Fe_4As_4F_2$ ($A$ = Rb, Cs) belong to extremely type-II superconductors. The superconducting magnetic hysteresis loops indicate strong magnetic flux pinning effect. Given the samples' dimensions with $e \times e \times h$ cm$^3$, the critical current density $J_c$ can be derived from the hysteresis loops using Bean critical state model [28], which gives $J_c(H) = 30 \times [M_+(H) - M_-(H)]/e$ (A cm$^{-2}$). Consequently, $J_c$ values of the RbCa$_2$Fe$_4$As$_4$F$_2$ sample are estimated to be $2.2 \times 10^5$ A cm$^{-2}$ at $H = 0$ and $8.0 \times 10^4$ A cm$^{-2}$ at $H = 10$ kOe, while $J_c$ values of the CsCa$_2$Fe$_4$As$_4$F$_2$ sample are approximately $5.2 \times 10^5$ A cm$^{-2}$ at $H = 0$ and $2.8 \times 10^5$ A cm$^{-2}$ at $H = 10$ kOe.

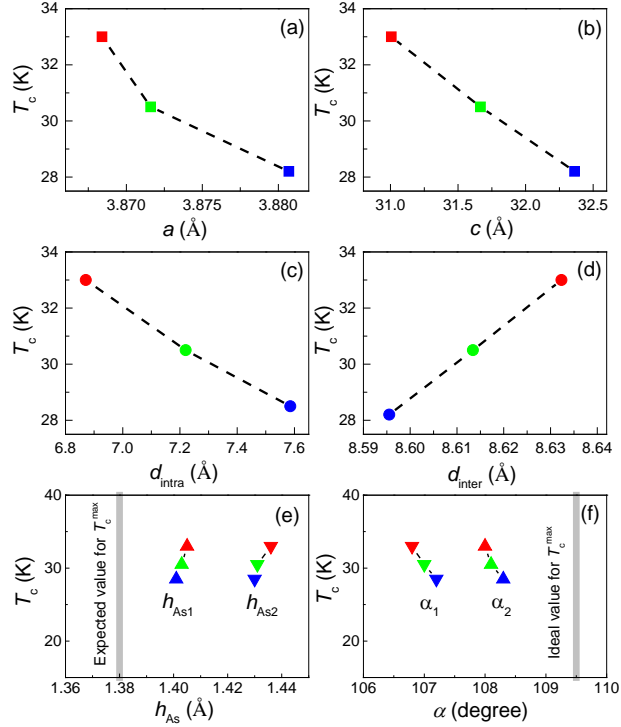

**Figure 4** Correlations between $T_c$ and structural parameters in $A$Ca$_2$Fe$_4$As$_4$F$_2$ ($A$ = K, Rb, Cs). The parameters include lattice constants $a$ (a) and $c$ (b), intra-Fe$_2$As$_2$-bilayer spacing (c), inter-Fe$_2$As$_2$-bilayer spacing (d), As height from the Fe plane, $h_{As}$ (e) and As–Fe–As bond angle, $\alpha$ (f). The thick gray lines mark the expected/ideal values of $h_{As}$ and $\alpha$ that generate maximum $T_c$.

As stated above, the hole doping level remains 25% for $A$Ca$_2$Fe$_4$As$_4$F$_2$ ($A$ = K, Rb, Cs) assuming ideal stoichiometric ratios. Therefore, it is meaningful to investigate the crystal-structure origin for the $T_c$ variations. Fig. 4 plots $T_c$ as functions of different structural parameters. One sees that $T_c$ inversely correlates with the lattice parameters $a$ and $c$. This is in sharp contrast to variations of $T_c^{max}$ in the optimally hole-doped ($Ae$, $Ak$)Fe$_2$As$_2$ systems ($T_c^{max}$ *increases* with the lattice parameters) [15-16]. Then, we consider the relevance of the spacing between Fe$_2$As$_2$ layers. While $T_c$ decreases with $d_{intra}$ (the spacing of Fe planes within the double Fe$_2$As$_2$ layers), in contrast, $T_c$ increases almost linearly with the spacing of Fe planes separated by the Ca$_2$F$_2$ layers, $d_{inter}$ [see Fig. 1(c)]. Given that the effective interlayer coupling increases with decreasing the spacing, the result suggests that either the intra-bilayer coupling enhances $T_c$, or, the inter-bilayer coupling suppresses $T_c$. If this is the case, $T_c$ values of IBS are determined not only by single Fe$_2$As$_2$ layer itself, as widely believed, but also by the coupling between different Fe$_2$As$_2$ layers, which calls for further theoretical investigations.

We note that the most commonly used parameters for the crystal-structure relationship with $T_c$ in IBS are the As–Fe–As bond angle, $\alpha$ [29,30] and the As height from the Fe plane, $h_{As}$ [31]. In the 12442-type structure, there are two different values for the above parameters due to the asymmetric $Fe_2As_2$ layers. The possible correlations with $T_c$ are shown in Fig. 4(e) & (f). As can be seen, both parameters deviate from the "ideal" values ($\alpha$ = 109.5° and $h_{As}$ = 1.38 Å) marked by thick gray lines. Contrarily, $T_c$ tends to increase when the parameters deviate more from the ideal values. The As height dependence resembles the $T_c$ variation in FeSe under high pressures [13].

In summary, we have synthesized and characterized new quinary iron-arsenide fluorides $RbCa_2Fe_4As_4F_2$ and $CsCa_2Fe_4As_4F_2$ with separate double $Fe_2As_2$ layers. The two compounds superconduct at 30.5 K and 28.2 K, respectively, due to self-doping effect. The $T_c$ trend violates the empirical rule of structural parameters. Instead, the interlayer coupling seems to be relevant to optimizing $T_c$. The success in obtaining new superconductors by incorporating large alkali-metal ions [18-22] is reminiscence of the recently discovered Cr-based superconductors $A_2Cr_3As_3$ [32-34], which suggests that the alkali-metal elements may have been overlooked in the exploration of superconducting materials. In the 12442-type IBS, alkali-metal cations seem essential, therefore, one may expect more 12442-type superconductors will be found by incorporation of alkali-metal cations in the near future.

**Experimental Section**

Polycrystalline samples of $ACa_2Fe_4As_4F_2$ ($A$ = Rb, Cs) were synthesized by solid-state reactions with the source materials Rb ingot (99.75%), Cs ingot (99.5%), Ca shot (99.5%), Fe powders (99.998%), As pieces (99.999%) and $CaF_2$ powders (99%). The intermediate products of CaAs, FeAs, and $Fe_2As$ were pre-synthesized with their corresponding elements at 1023 K for 12 hours in evacuated quartz tube, respectively. Then $Rb_{1.03}Fe_2As_2$ and $Cs_{1.03}Fe_2As_2$ (3% excess in order to compensate the loss of alkali metal) were prepared by reacting Rb and Cs with FeAs at 923 K and 873 K for 10 hours as another two intermediate products. Moreover, $CaF_2$ was heated to 873 K for 24 hours in air to remove adsorbed water. After that, $RbFe_2As_2$ (or $CsFe_2As_2$), CaAs, $CaF_2$ and $Fe_2As$ were mixed homogeneously in the stoichiometric ratio and pressed into pellets. The pellets were loaded in an alumina tube with a cover. The sample-loaded alumina tube was sealed in a Ta tube which was jacketed with an evacuated quartz ampoule. Subsequently, the sample was heated to 1203 K in 5 hours, holding for 36 hours. The final products were stable in air. All the operations above were carried out in an argon-filled glovebox with the water and oxygen content below 1 ppm.

Powder X-ray diffraction was carried out at room temperature on a PANalytical x-ray diffractometer (Empyrean Series 2) with a monochromatic CuK$\alpha_1$ radiation. The electrical transport measurements were carried out by a standard four-terminal method on a Cryogenic Limited Vibrating Sample Magnetometer (VSM) equipped with a Keithley 2400 digital sourcemeter and a Keithley 2182 nanovoltmeter. The dc magnetizations were measured on a Quantum Design Magnetic Property Measurement System (MPMS-XL5), and the samples were cut into a rectangular shape so as to obtain the demagnetization factor easily.

**References**

1. Kamihara Y, Watanabe T, Hirano M and Hosono H. Iron-based layered superconductor La[$O_{1-x}F_x$]FeAs ($x$ = 0.05-0.12) with $T_c$ = 26 K. J Am Chem Soc, 2008, 130: 3296-3297
2. Chen XH, Wu T, Wu G, Liu RH, Chen H and Fang DF. Superconductivity at 43 K in SmFeAsO$_{1-x}$F$_x$. Nature, 2008, 453: 761-762
3. Rotter M, Tegel M and Johrendt D. Superconductivity at 38 K in the Iron Arsenide (Ba$_{1-x}$K$_x$)Fe$_2$As$_2$. Phys Rev Lett, 2008, 101: 107006-107009
4. Wang XC, Liu QQ, Lv YX, *et al*. The superconductivity at 18 K in LiFeAs system. Solid State Commun, 2008, 148: 538–540
5. Hsu FC, Luo JY, Yeh KW, *et al*. Superconductivity in the PbO-type structure α-FeSe, Proc Natl Acad Sci (USA), 2008, 105, 14262-14264
6. Guo JG, Jin SF, Wang G. Superconductivity in the iron selenide K$_x$Fe$_2$Se$_2$ (0≤x≤1.0). Phys. Rev. B, 2010, 82: 180520-180523
7. Katayama N, Kudo K, Onari S, *et al*. Superconductivity in Ca$_{1-x}$La$_x$FeAs$_2$: A Novel 112-Type Iron Pnictide with Arsenic Zigzag Bonds. J Phys Soc Jpn, 2013, 82: 123702
8. Zhu XY, Han F, Mu G, *et al.* Sr$_3$Sc$_2$Fe$_2$As$_2$O$_5$ as a possible parent compound for FeAs-based superconductors. Phys. Rev. B, 2009, 79: 024516-024520
9. Ogino H, Matsumura Y, Katsura Y, *et al*. Superconductivity at 17 K in (Fe$_2$P$_2$)(Sr$_4$Sc$_2$O$_6$): a new superconducting layered pnictide oxide with a thick perovskite oxide layer. Supercond Sci Technol, 2009, 22: 075008-075011
10. Kakiya S, Kudo K, Nishikubo Y, *et al*. Superconductivity at 38 K in iron-based compound with platinum-arsenide layers Ca$_{10}$(Pt$_4$As$_8$) (Fe$_{2-x}$Pt$_x$As$_2$)$_5$. J Phys Soc Jpn, 2011, 80: 093704
11. Ni N, Allred J M, Chan B C and Cava R J. High $T_c$ electron doped Ca$_{10}$(Pt$_3$As$_8$)(Fe$_2$As$_2$)$_5$ and Ca$_{10}$(Pt$_4$As$_8$)(Fe$_2$As$_2$)$_5$ superconductors with skutterudite intermediary layers. Proc Natl Acad Sci USA, 2011, 108: E1019-E1026
12. Sun YL, Jiang H, Zhai HF, *et al.* Ba$_2$Ti$_2$Fe$_2$As$_4$O: A New Superconductor Containing Fe$_2$As$_2$ Layers and Ti$_2$O Sheets. J Am Chem Soc, 2012, 134: 12893-12896
13. Lu XF, Wang NZ, Wu H, *et al*. Coexistence of superconductivity and antiferromagnetism in (Li$_{0.8}$Fe$_{0.2}$)OHFeSe. Nat Mater, 2015, 14: 325-329
14. Johnston DC. The puzzle of high temperature superconductivity in layered iron pnictides and chalcogenides. Adv Phys, 2010, 59: 803-1061
15. Chen XH, Dai, PC, Feng DL, Xiang T and Zhang FC. Iron-based high transition temperature superconductors. Natl Sci Rev, 2014, 1: 371-395
16. Luo, XG and Chen XH. Crystal structure and phase diagrams of iron-based superconductors. Sci China Mater, 2015, 58:77-89
17. Jiang H, Sun YL, Xu ZA, and Cao GH. Crystal chemistry and structural design of iron-based superconductors. Chin Phys B, 2013, 22: 087410-087420
18. Iyo A, Kawashima K, Kinjo T, *et al*. New-Structure-Type Fe-Based Superconductors: Ca*A*Fe4As4 (*A* = K, Rb, Cs) and Sr*A*Fe4As4 (*A* = Rb, Cs). J Am Chem Soc, 2016, 138: 3410-3415
19. Liu Y, Liu YB, Tang ZT, *et al*. Superconductivity and ferromagnetism in hole-doped RbEuFe$_4$As$_4$. Phys. Rev. B, 2016, 93: 214503
20. Liu Y, Liu YB, Chen Q, *et al.* A new ferromagnetic superconductor: CsEuFe$_4$As$_4$, Sci Bull, 2016, 61: 1213-1220
21. Kawashima K, Kinjo T, Nishio T, *et al*. Superconductivity in Fe-Based Compound Eu*A*Fe$_4$As$_4$ (*A* =

Rb and Cs). J Phys Soc Jpn, 2016, 85: 064710

22. Wang ZC, He CY, Wu SQ, *et al.* Superconductivity in KCa$_2$Fe$_4$As$_4$F$_2$ with Separate Double Fe$_2$As$_2$ Layers. J Am Chem Soc, 2016, 138: 7856-7859
23. Izumi F and Momma K. Three-dimensional visualization in powder diffraction. Solid State Phenomena, 2007, 130: 15-20
24. Bukowski Z, Weyeneth S, Puzniak R, Bulk superconductivity at 2.6 K in undoped RbFe$_2$As$_2$. Physica C, 2010, 470: S328–S329
25. Matsuishi S, Inoue Y, Nomura T, *et al.* Superconductivity Induced by Co-Doping in Quaternary Fluoroarsenide CaFeAsF. J Am Chem Soc, 2008, 130: 14428-14429
26. Noack M and Schuster H-U. New Ternary Compounds of Cesium and Elements of the 8$^{th}$ Transition Metal Group and the 5$^{th}$ Main Group. Z anorg allg Chem, 1994, 620: 1777-1780
27. Wen HH, Mu G, Fang L, *et al.* Superconductivity at 25 K in hole-doped (La$_{1-x}$Sr$_x$)OFeAs. Europhys Lett, 2008, 82: 17009-17013
28. Bean C. Magnetization of high-field superconductors. Rev Mod Phys, 1964, 36: 31-39
29. Lee CH, Iyo A, Eisaki H, *et al.* Effect of Structural Parameters on Superconductivity in Fluorine-Free LnFeAsO$_{1-y}$ (Ln = La, Nd). J Phys Soc Jpn, 2008, 77: 083704-083707
30. Zhao J, Huang Q, Cruz C, *et al*. Structural and magnetic phase diagram of CeFeAsO$_{1-x}$F$_x$ and its relation to high-temperature superconductivity. Nat Mater, 2008, 7: 953-959
31. Mizuguchi Y, Hara Y, Deguchi Y, *et al.* Anion height dependence of $T_c$ for the Fe-based superconductor. Supercond Sci Technol, 2010, 23: 054013-054017
32. Bao JK, Liu JY, Tang TZ, *et al*. Superconductivity in quasi-one-dimensional K$_2$Cr$_3$As$_3$ with significant electron correlations. Phys Rev X, 2015, 5: 011013
33. Tang ZT, Bao JK, Liu Y, *et al*. Unconventional Superconductivity in quasi-one-dimensional Rb$_2$Cr$_3$As$_3$. Phys Rev B, 2015, 91: 020506
34. Tang ZT, Bao JK, Wang Z, *et al*. Superconductivity in quasi-one-dimensional Cs$_2$Cr$_3$As$_3$ with large interchain distance. Sci China Mater, 2015, 58: 16-20

**Acknowledgements** This work was supported by the Natural Science Foundation of China (Nos. 90922002 and 11190023) and the National Key Research and Development Program of China (No. 2016YFA0300202)

**Author contributions** Cao GH designed the experiment, discussed the result and wrote the paper with Wang ZC. Wang ZC synthesized the samples together with He CY. The structural characterizations and physical property measurements were performed with assistance from Tang ZT and Wu SQ.

**Conflict of Interest** The authors declare that they have no conflict of interest.